# A simple model of dust extinction in gamma-ray burst host galaxies

N. A. Rakotondrainibe[1,★], V. Buat[1], D. Turpin[2], D. Dornic[3], E. Le Floc'h[2], S. D. Vergani[4], and S. Basa[1]

[1] Aix Marseille Univ., CNRS, CNES, LAM, Marseille, France
[2] Université Paris-Saclay, Université Paris Cité, CEA, CNRS, AIM, 91191 Gif-sur-Yvette, France
[3] Aix Marseille Univ., CNRS/IN2P3, CPPM, Marseille, France
[4] GEPI, Observatoire de Paris, Université PSL, CNRS, 5 Place Jules Janssen, 92190 Meudon, France

**ABSTRACT**

*Context.* Gamma-ray burst (GRB) afterglows are powerful probes for studying the different properties of their host galaxies (e.g., the interstellar dust) at all redshifts. By fitting their spectral energy distribution (SED) over a large range of wavelengths, we can gain direct insights into the properties of the interstellar dust by studying the extinction curves. Unlike the dust extinction templates, such as those of the average Milky Way (MW) or the Small and Large Magellanic Cloud (SMC and LMC), the extinction curves of galaxies outside the Local Group exhibit deviation from these laws. Altogether, X-ray and gamma-ray satellites as well as ground-based telescopes, such as *Neil Gehrels Swift* Observatory (*Swift*) and Gamma-Ray Optical and Near-Infrared Detector (GROND), provide measurements of the afterglows from the X-ray to the NIR, which can be used to extract information on dust extinction curves along their lines of sight (LoS). The study presented in this paper undertakes such a photometric study, comprising a preparatory work for the SVOM mission and its ground-based follow-up telescope COLIBRI.
*Aims.* We propose a simple approach to parameterize the dust extinction curve of GRB host galaxies. The model used in this analysis is based on a power law form with the addition of a Loretzian-like Drude profile with two parameters: the extinction slope, $\gamma$, and the 2175 Å bump amplitude, $E_b$.
*Methods.* Using the $g'r'i'z'JHK_s$ GROND filter bands, we tested our dust extinction model and explored the parameter space in extinction and redshift by fitting SEDs of simplified simulations of GRB afterglow spectra based on different extinction curve templates. From a final sample of 10 real *Swift*/GROND extinguished GRBs, we determined the quantities of the dust extinction in their host and measured their extinction curves.
*Results.* We find that our derived extinction curves are in agreement with the spectroscopic measurements reported for four GRBs in the literature. We compared four other GRBs to the results of photometric studies where fixed laws were used to fit their data. We additionally derived two new GRB extinction curves. The measured average extinction curve is given by a slope of $\gamma = 1.051 \pm 0.129$ and $E_b = 0.070 \pm 0.036$, which is equivalent to a quasi-featureless in-between SMC-LMC template. This is consistent with previous studies aimed at deriving the dust host galaxy extinction where we expect that small dust grains dominate in GRB environment, yielding a steeper curve than the mean MW extinction curve.

## 1. Introduction

Gamma-ray burst (GRB) jets are believed to be generated by compact central engines (stellar mass black holes or highly magnetized neutrons stars) rapidly accreting surrounding materials coming either from the collapse of massive stars (long GRBs, Mészáros 2006) or the merger of two compact objects (including at least one neutron star) in a close binary system (short GRBs, D'Avanzo 2015). Two main phases can be observed during the GRB phenomenon. Detected at high energies, the prompt phase is the short and intense emission of hard X-rays and gamma rays, lasting from a few milliseconds to several tens of seconds. It is followed by the afterglow that is powered by the collisions between the ultra-relativistic jets and the surrounding interstellar medium in the vicinity of the explosion site. Particle acceleration mechanisms at the shock discontinuity make the ejecta radiate in a broad range of wavelengths through synchrotron and inverse Compton radiative processes (Granot & Sari 2002).

Because of their high luminosities, they are visible up to cosmological distances. The rest-frame ultraviolet (UV) absorption of the luminous GRB intrinsic spectra makes them effective probes for studying the dust properties in their surrounding environment. Indeed, the extinction caused by the absorption and scattering in the host galaxy dust induce significant deviations from their power law shaped spectra. The study of GRB lines of sight (LoS) within their host galaxies has been shown to be an excellent way to measure the extinction from dust in galaxies beyond the local universe (e.g., Galama & Wijers 2001; Kann et al. 2006; Schady 2017). Understanding the nature of the cosmic dust is important for inferring the properties of the interstellar medium (ISM) when the Universe was younger and galaxies less evolved than in the present time. Other phenomena have also been used to study dust extinction such as reddened quasi-stellar objects (QSOs) but are less robust compared to the use of GRBs (Richards et al. 2003; Jiang et al. 2011; Zafar et al. 2015). Different theoretical models also predict the progenitors of dust at high redshifts specifically. At $z > 4-6$, the extinction caused by dust produced by supernovae (SNe) (Todini & Ferrara 2001) is expected, yet it is rarely observed and remains a controversial topic of debate (Perley et al. 2010; Stratta et al. 2011; Jang et al. 2011).

The most common method for the analysis of GRB host dust extinction consists of fitting the broadband X-ray to near-infrared (NIR) spectral energy distribution (SED) of GRB afterglows. With the quasi-unattenuated X-ray spectrum at ≳2 keV, it allows for the intrinsic afterglow spectral index to be accurately determined. Thus, precise measurements of the imprints from dust extinction can be obtained. The UV to NIR SEDs are fitted using templates of extinction curves. The most frequently adopted ones are the fixed templates taken from the average Milky Way (MW), the Large (LMC) and Small Magellanic Clouds (SMC) LoS such as in the study by Pei (1992). The main differences between each model are the strength of the rest-frame 2175 Å bump feature and their overall slope. The MW extinction curve shows a strong bump, which is less pronounced in the LMC and absent of the SMC curve (Gordon et al. 2003). Its strength is anti-correlated with the steepness of the three extinction curves at UV wavelengths. The main origin of this 2175 Å absorption remains however not fully resolved after its first detection (Stecher 1965, 1969). It is believed to be caused by aromatic carbonaceous (graphitic) materials, likely a cosmic mixture of polycyclic aromatic hydrocarbon (PAH) molecules (Wickramasinghe & Guillaume 1965; Joblin et al. 1992; Li & Draine 2001; Draine 2003). Generally, the SMC featureless extinction law is the model which best fits the GRB afterglows (Kann et al. 2006, 2010; Greiner et al. 2011; Japelj et al. 2015). The presence of small grains could be the cause of the steep extinction curve in GRB hosts, which may be the result of dust being destroyed by the extreme conditions in the ISM (Reach et al. 2000). It is important to recall that there is no reason to believe that the known MW, LMC, SMC extinction curves would represent those beyond the Local Group, as they display a broad range of ages, luminosities, metallicities, and star formation processes (Hou et al. 2016, 2017).

★ Corresponding author; ny-avo.rakotondrainibe@lam.fr

Multiple parameterized laws have been used to measure the extinction curves of GRB host galaxies (e.g., Fitzpatrick & Massa 1990 hereafter FM90 and its modified parameterizations, Li et al. 2008). These methods can lead to higher degeneracies in the fitting process of GRB SEDs as their flexibility depends on the quality and quantity of UV-to-NIR data due to their high number of parameters. Therefore, they are less suitable for photometric datasets where the poor resolution is insufficient to constrain these parameters strongly. Ultimately, only a few number of extinction curves have been measured up to now and they differ only slightly or greatly from the fixed templates (Zafar et al. 2011, 2018b). This confirms the non-universality of the extinction law of GRB host galaxies.

We are now in the era of rapid-response observations of GRBs. Multiwavelength observations are essential in GRBs and other transient phenomena studies. The importance of a rapid follow-up is crucial as the analysis of GRB extinction curves may be impacted by the quick fade of GRB afterglows. Early-time observations taken with the onboard *Neil Gehrels Swift* Observatory (*Swift*) (Gehrels et al. 2004) instruments enable the study of GRBs up to the epoch of reionization. In addition, optical-to-NIR data from ground-based telescopes such as the seven-band filter Gamma-Ray Optical and Near-Infrared Detector (GROND, Greiner et al. 2008), have a high enough sensitivity and wide-band coverage for studying the dust properties of GRB host galaxies.

The main goal of this paper is to provide a simple extinction model able to retrieve the main properties of the extinction laws identified in GRB host galaxies in this context of photometric observations combined with X-ray data. We propose a simple extinction curve, based on a power law model with the extinction slope as the main parameter. This type of approach has already been investigated in the context of the extinction in the Galaxy (Divan 1954; Nandy & Royal Observatory, E. 1964; Zagury 2012) in the visible/NIR and in GRB host galaxy (GRB 020813, Savaglio & Fall 2004).

In Sect. 2, we introduce our dust extinction model and the extinction curves that were used for the simulation of 'perfect' GRB afterglows. This is meant to test and to validate our model on idealized cases. The steps from the simulation of the afterglow populations to the results of the fits using our model are detailed in Sect. 3. The application of our model on a selection of GRBs observed by *Swift*/GROND is described in Sect. 4. We discuss our methodology and present our conclusions in Sects. 5 and 6.

The errors throughout this paper on the level of $1\sigma$ unless specified otherwise.

## 2. Modeling the dust extinction in GRB host galaxies

In this section, we present the selection of observed dust extinction curves that we seek to parameterize with our simple model.

### 2.1. Empirical dust extinction models in the Universe

The analytic representations of the extinction curves in the MW, SMC and LMC are the most commonly used models of extinction for studying the GRB host dust extinction. To model them, we chose to employ the parameterization of Pei (1992). The use of these average extinction curves instead of the Gordon et al. (2003) parameterized SMC and LMC or the different MW laws of Fitzpatrick & Massa (1986, hereafter FM86) is deliberate as the slight differences in their shapes justify the use of these models. Nevertheless, even within our own Galaxy, observations have revealed a significant diversity of extinction laws (Fitzpatrick & Massa 2007 hereafter FM07). Most of the observed Galactic sightlines are consistent with the variation in the total-to-selective extinction coefficient $R_V$. This parameter alone has been used to represent the extinction laws at optical through UV wavelengths (Cardelli et al. 1989). To investigate these variations, we considered extinction curves derived from observations of different MW sightlines. Five sightlines spanning from $R_V = 2.3$ to 7.5 were chosen to explore curve behaviors from a nearly flat with a minimal 2175 Å absorption to the steep extinction curve found along the sightline exhibited by HD 211021 (Larson et al. 2000). The extinction curve observed by Maiolino et al. (2004) in the most distant broad absorption line quasar (BAL QSO) at $z = 6.2$ was also investigated in our approach. This unusual model of extinction presents a flat region between 1700 and 3300 Å and a rapid increase below 1700 Å. Another template integrated in this work is the average QSO extinction curve (Zafar et al. 2015). Indeed, adding an even steeper model than in the MW towards HD 210121 increases the range to test our simpler model.

### 2.2. A simpler analytical approach for dust extinction modeling

On a general note, GRB extinction curves typically show minimal deviation from an SMC-like shape (Schady et al. 2012a; Zafar et al. 2011, 2018b). which exhibits a power law behavior. In contrast, the 2175 Å absorption feature is rarely observed in GRB LoS (Kann et al. 2006; Greiner et al. 2008; Japelj et al. 2015; Schady 2017; Heintz et al. 2023). By assuming that the general UV-to-NIR extinction can be described by a power law,

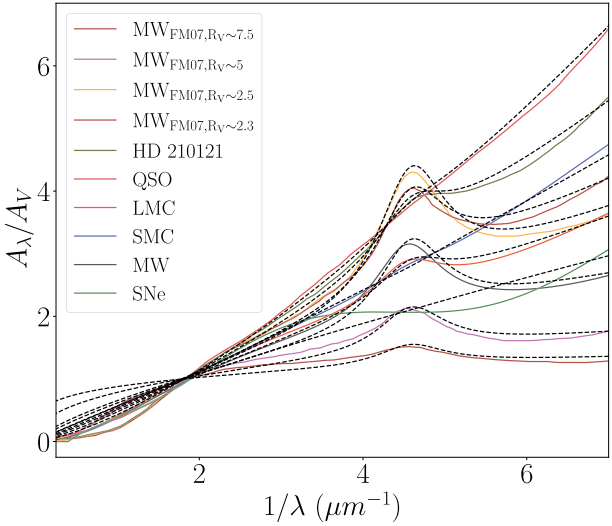

**Fig. 1.** Selection of extinction curves used for the creation of mock populations of "perfect" GRB afterglows (solid lines) and best fits of the power law extinction curve (dashed lines).

we propose constructing it by using a normalized curve to the $V$-band described by

$$A_\lambda = A_V \left(\frac{5500\,\text{Å}}{\lambda}\right)^\gamma \quad (1)$$

with $A_V$ as the amount of extinction along the LoS of the host galaxy and $\gamma$ the extinction slope.

Nonetheless, even considering the small detection rate of the 2175 Å absorption feature in GRB host extinction curves (Elíasdóttir et al. 2009; Krühler et al. 2008; Perley et al. 2011; Zafar et al. 2012), we still account for it in our model. The addition of a Lorentzian-like Drude profile is commonly used to characterize this feature observed in MW-like extinction curve (Fitzpatrick 1999 hereafter FM99). Thus, Eq. (1) becomes

$$A_\lambda = A_V \left[\left(\frac{5500\,\text{Å}}{\lambda}\right)^\gamma + D(E_b, \lambda)\right] \quad (2)$$

where

$$D(E_b, \lambda) = \frac{E_b (\lambda \Delta\lambda)^2}{(\lambda^2 - \lambda_0^2) + (\lambda \Delta\lambda)^2} \quad (3)$$

with $E_b$ as the maximum amplitude of the bump above the UV linear extinction, while $\lambda_0$ and $\Delta\lambda$ are the central wavelength of the bump and its width, respectively. As the central wavelength of the 2175 Å bump does not present any significant variations (Mathis 1994; Rouleau et al. 1997), we fix $\lambda_0 = 2175$ Å. Given the low spectral resolution in a photometric framework compared to spectroscopy, we also fix the bump width to the median MW value of $\Delta\lambda = 470$ Å (FM99).

As shown in Fig. 1, fits of this model using the Levenberg–Marquardt algorithm with the resulting values of $\gamma$ and $E_b$ are able to reproduce the main shapes of the different extinction curve templates used in our analysis. The fact that the SNe extinction curve is unique, due to its plateau feature, makes it impossible to reproduce with our template. Also, given the rarity of this case, we can opt to omit it from further analysis. The corresponding values for each fit are reported in Table 1 and are indicated with the subscript "$_{\text{ref}}$".

## 3. Testing on simulated GRB afterglows

In this section, our objective is to test whether the parameters of our simple dust extinction model described in Sect. 2.2 can be constrained by a set of photometric data, assuming that the intrinsic UV-to-NIR emission of the afterglow is known. The determination of this afterglow spectrum relies on X-ray measurements combined with optical data. We introduce this analysis in Sect. 4 with the observed GRB data.

### 3.1. The GRB afterglow simulation

In theory, GRBs are produced by the fireball shock model, while GRB afterglows originate from synchrotron radiation caused by the interaction between the ultra-relativistic jet and the ISM (e.g., Piran 1999; Granot & Sari 2002). This synchrotron spectrum consists of multiple power laws connected at typical break frequencies that evolve towards redder wavelengths as a function of time. In this framework, the UV-to-NIR wavelengths are generally set on the same power law branch of the synchrotron spectrum. To determine the intrinsic behaviors of the parameters of our model, we aim to only characterize the dust extinction in these wavelengths. Thus, our model is investigated on idealized cases of a power law afterglow by assuming that its properties are known. The case where it is to be determined along the dust extinction properties is explored in the next section with X-ray-to-NIR data.

Our GRB afterglow spectra were then created using a simple frequency-dependent power law, which is expressed as:

$$F_\nu = F_0 \times \nu^{-\beta} \quad (4)$$

where $F_0$ is the flux normalization, $\nu$ is the radiation frequency and $\beta$ the intrinsic spectral slope. This intrinsic flux is extinguished due to dust in the host galaxy. The extinction is expressed as a function of wavelength by the extinction curve $A_\lambda$ thus the extinguished flux is given by

$$F_\nu^{\text{obs}} = F_\nu \times 10^{-0.4\,A_\lambda}. \quad (5)$$

The spectral shapes of the simulated afterglows were constrained by fixing the flux normalization and the slope to constant values at $F_0 = 1$ mJy and $\beta = 0.66$, respectively, as defined in Eq. (4). The intrinsic slope was selected based on the distribution of $\beta$ values reported in Kann et al. (2010). To investigate our model's behavior in response to the variability of the slope that we can observe, we also did our analysis using different values of this slope (at $3\sigma$ errors of $\beta = 0.66$). Because of the consistency of our results, only the ones with $\beta = 0.66$ are reported. By fixing these parameters, this allowed us to focus solely on identifying the impact of the extinction curves on given standard GRB spectra without the degeneracies that can be brought from the determination of the intrinsic spectrum.

The GRB populations were created using the nine different dust extinction templates introduced in Sect. 2.1. A grid of GRB spectra was then built considering different values of the parameters ($A_V, z$), ranging from $0.5 < z < 6$ and $0.1 < A_V < 2.4$ mag. We chose these ranges based on results from analysis of unbiased samples of GRBs (e.g., Covino et al. 2013), while also ensuring that a large fraction of the parameter space was covered. This led us to 3519 bursts to test our simplified model of extinction.

We extracted the optical-to-NIR afterglow flux values at each central wavelength of the GROND $g'r'i'z'JHK_s$ bands and added an uncertainty of 15% for each. For this simulation part, no physical or instrumental aspects such as the Galactic

**Table 1.** Values of the extinction slope $\gamma$ and the 2175 Å bump amplitude $E_b$ for the studied extinction laws.

| Param. $A_\lambda/A_V$ | MW | MW$_{FM07, R_V \sim 2.5}$ | MW$_{FM07, R_V \sim 5}$ | MW$_{FM07, R_V \sim 7.5}$ |
|---|---|---|---|---|
| $\gamma_{ref}$ | 0.714 ± 0.027 | 0.961 ± 0.034 | 0.403 ± 0.035 | 0.218 ± 0.034 |
| $E_{b,ref}$ | 1.295 ± 0.014 | 1.959 ± 0.024 | 0.696 ± 0.013 | 0.328 ± 0.010 |
| $\gamma_{fit}$ | 0.668 ± 0.085 | 0.856 ± 0.141 | 0.421 ± 0.155 | 0.199 ± 0.132 |
| $E_{b,fit}$ | 1.366 ± 0.449 | 1.866 ± 0.667 | 0.644 ± 0.506 | 0.364 ± 0.567 |
| | MW$_{FM07, R_V \sim 2.3}$ | HD 211021 | LMC | SMC | QSO |
| $\gamma_{ref}$ | 1.051 ± 0.025 | 1.251 ± 0.016 | 0.944 ± 0.015 | 1.129 ± 0.007 | 1.403 ± 0.018 |
| $E_{b,ref}$ | 1.401 ± 0.017 | 0.731 ± 0.012 | 0.516 ± 0.009 | 0.039 ± 0.004 | 0.172 ± 0.257 |
| $\gamma_{fit}$ | 1.044 ± 0.112 | 1.264 ± 0.138 | 0.938 ± 0.046 | 1.193 ± 0.047 | 1.386 ± 0.009 |
| $E_{b,fit}$ | 1.303 ± 0.653 | 0.665 ± 0.388 | 0.469 ± 0.232 | 0.069 ± 0.267 | 0.060 ± 0.257 |

**Notes.** $_{ref}$ corresponds to values from direct fits of the nine extinction curves $A_\lambda/A_V$. $_{fit}$ corresponds to values from fits of the GRB simulated afterglows based on the nine extinction laws at $z > 1.5$ and $A_{V,host} > 0.3$ mag.

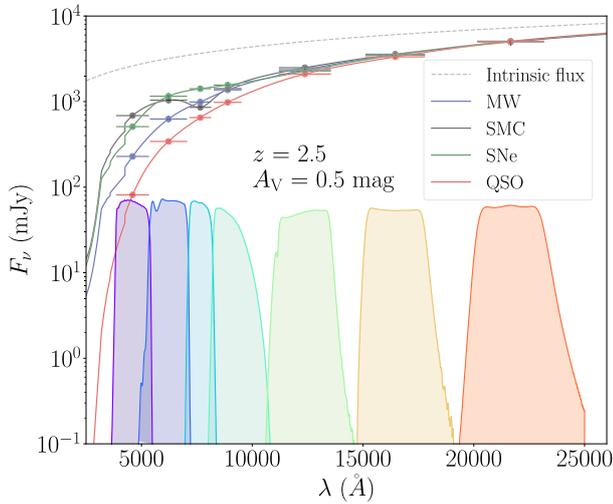

**Fig. 2.** Impact of the SMC, MW, SNe and QSO dust extinction curve on the observed GRB afterglow optical-to-NIR spectrum at $z = 2.5$ with $A_V = 0.5$ mag. The grey dotted line represents the GRB afterglow spectrum with no absorption. We also show the GROND seven-band filters global throughput and the derived photometric points of the spectra.

foreground reddening, the soft X-ray absorption, or the instrument sensitivity were considered. This approach was designed to establish an initial application of our proposed simplified extinction law under idealized conditions. However, the Ly$\alpha$ absorption arising from the intergalactic medium (IGM) clouds (Meiksin 2006) was taken into account. Examples of simulated extinguished GRB optical-to-NIR afterglow spectra and their associated photometric points in the GROND passband filters are shown in Fig. 2.

### 3.2. Parameters estimation using the 'simplified' dust extinction model

By assuming that the redshift of the GRB is known, we went on to pursue to estimate the three remaining parameters, $\gamma$, $A_V$, $E_b$, of our simple extinction model. We used the Markov chain Monte Carlo (MCMC) Python implementation, emcee (Foreman-Mackey et al. 2013). These parameters were randomly drawn within the prior intervals defined as $0 < \gamma < 3$, $0 < A_V < 5$ mag and $-1 < E_b < 5$. Large intervals of the parameter space were used to allow greater flexibility in constraining the estimation process. We employed a total of nine chains to explore the parameter space efficiently and allowed them to evolve over 750 steps. The burn-in phase was set to 30% of the step number to ensure the convergence before collecting samples for posterior analysis.

In the following, the derived value of the amount of $V$-band extinction in the host is denoted as $A_V$ and the fixed value used for the simulation is $A_{V,host}$.

#### 3.2.1. Estimation of the extinction, $A_V$

Not only is the determination of the extinction curve important in building a better understanding of the dust properties in galaxies, but also its normalization – in our case, in the $V$-band. It is shown in Fig. 3 that this quantity is well retrieved using our model. However, we do notice a trend of dispersion particularly among spectra based on the MW-like extinction curve. An underestimation of values of $A_V$ at redshift $z < 1.5$ is observed with our model for observations without UV coverage. This underestimation is more pronounced with higher values of $A_{V,host}$. When a greater bump lies in the bluer bands, a fit with our model will consider the extinction slope as the deviation of the intrinsic afterglow from the extinction $A_V$. Such extreme cases of low redshift and high extinction have never been observed outside the Local Group. This trend is not visible for featureless laws like the SMC or the QSO.

#### 3.2.2. Estimation of the extinction slope, $\gamma$

In contrast to the different extinction laws dependent on the dust properties along the LoS, the shape of the power law extinction model explored in this study is exclusively tied to the parameter $\gamma$. The evolution of the values of the extinction slope as a function of the redshift is shown in Fig. 4. We highlight here the results from the populations of GRB spectra using different templates of extinction laws at a moderate $A_{V,host} = 0.5$ mag for reference. In each case, a discernible trend in the values of the extinction slope is observed at $z < 1.5$. For low redshifts, our restricted photometric coverage prevents us from accurately exploring the impact of the dust extinction at UV wavelengths and effectively constrains $\gamma$ within these ranges. Therefore, we obtain much weaker constraints on the dust extinction law. It has an even greater effect on the constraints on models presenting the distinct spectral features 2175 Å absorption.

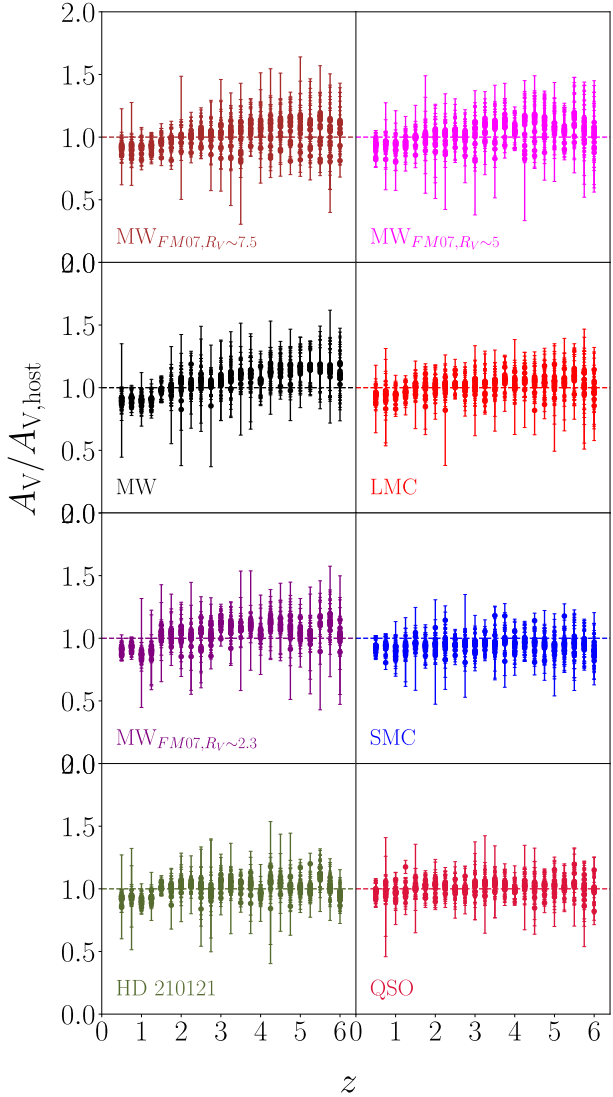

**Fig. 3.** Accuracy of the estimation of $A_V$ expressed as the evolution of the ratio $A_V/A_{V,\text{host}}$ as a function of redshift, $z$.

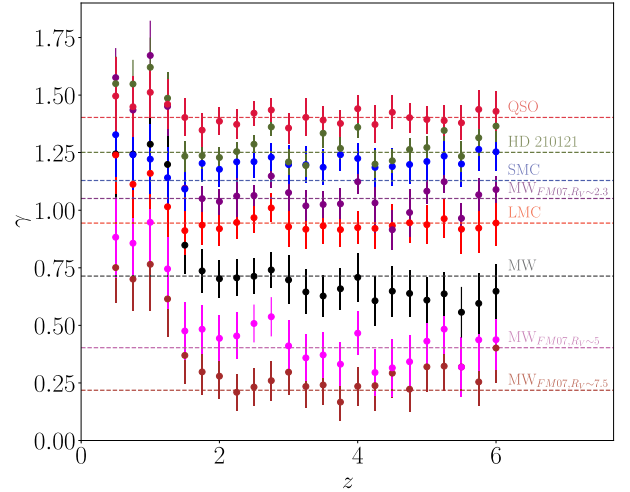

**Fig. 4.** Evolution of the extinction slope $\gamma$ estimation as a function of redshift, $z$, at $A_{V,\text{host}} = 0.5$ mag. Dashed lines correspond to the reference values of $\gamma$ for each law (Table 1).

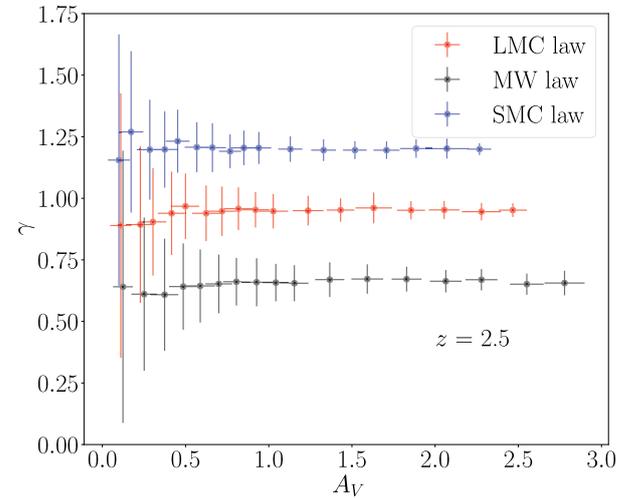

**Fig. 5.** Evolution of the extinction slope $\gamma$ estimation as a function of the extinction $A_V$ for $z = 2.5$ for the three typical LMC, MW and SMC extinction laws.

Exploring the variation in $\gamma$ with respect to $A_V$ also sheds light on the difficulties associated with studying the host galaxy at low extinction. By focusing on the three mean laws of MW, LMC, and SMC, it is revealed in Fig. 5 that the differentiation between these typical models becomes clear only at substantial values of the extinction, primarily, $A_V > 0.3$ mag. This is due to the large uncertainties on the estimation of the slope $\gamma$.

### 3.2.3. Estimation of the 2175 Å bump amplitude, $E_b$

As expected, the challenge of clearly observing the 2175 Å bump increases as the extinction in the host galaxy diminishes. Different studies have suggested that this feature is mostly observed in highly extinguished afterglows (Schady et al. 2012a; Covino et al. 2013; Zafar et al. 2018b). The ability to detect this feature can also be expected only under specific conditions, especially in a photometric study. The redshift affects not only the afterglow spectrum but also the bump which is shifted and widened when $z$ increases. The ability to detect and measure the bump amplitude $E_b$ is therefore dependent on the wavelength band coverage of the observations. Figure 6 shows its evolution at a redshift where the bump is at the central wavelength of a filter band (the $i$ band here). It is evident that at lower extinction, the three laws are nearly indistinguishable in terms of the bump amplitude.

### 3.3. Summary

We fit the GRB populations based on a selection of extinction curves using our simple model of extinction. Values of the extinction slope $\gamma$ and the 2175 Å bump amplitude $E_b$ for the nine templates of extinction in this idealized study are listed in Table 1 (indicated subscript "fit"). We have shown that our simple model of extinction curve can recover the typical parameterizations of extinction in GRB host galaxies. However, carrying out this study with simulated afterglows led to some necessary conditions to retrieve the extinction parameters from photometric observations. The 2175 Å bump is not measurable at redshift $z < 1.5$ due to the lack of UV coverage. The amount of extinction is also crucial for distinguishing between the different extinction laws. In this work, it is found that it is only possible to clearly distinguish this difference at a substantial extinction of $A_V > 0.3$ mag. We then limit our analysis on GRB data within the ranges of these two parameters at redshift $z > 1.5$ and an extinction of $A_V > 0.3$ mag (Sect. 4). Also, fixing different parameters, namely the width of the bump and its central wavelength, could lead to degeneracies by not having flexibility in the overall fits. However, this can be addressed through spectroscopy or multi-broadband photometric observations by using more refined parameterizations that can constrain these quantities.

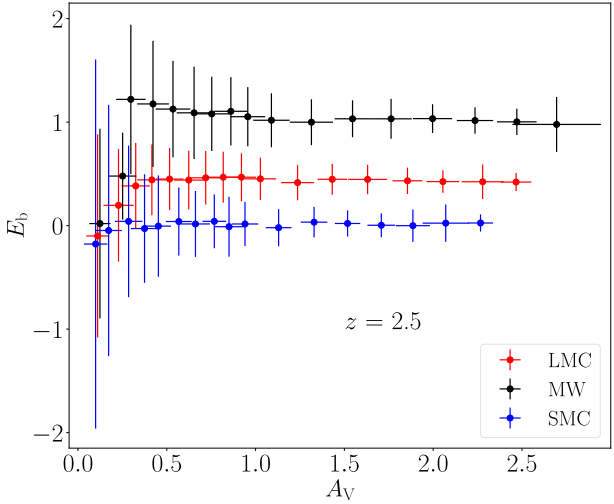

**Fig. 6.** Evolution of the estimations of $E_b$ as a function of $A_V$ for $z = 2.5$ for the three typical LMC, MW and SMC extinction laws.

## 4. Application to real *Swift*/GROND GRBs

After investigating our dust extinction model (Sect. 3) with simulations by assuming that the intrinsic afterglow emission is determined, we now apply our model to a sample of GRBs observed by *Swift*/GROND. With X-ray-to-NIR data, the intrinsic continuum properties are determined in addition to our model's 3 parameters which describe the dust extinction properties of GRB host galaxies.

### 4.1. GRB selection criteria

Starting with the table of 498 GRB sources[1] with known spectroscopic or photometric redshift, we have selected our samples by applying several criteria. Following the outcomes of our studies (Sect. 3.2), we chose GRBs with $z > 1.5$. Additionally, we chose GRBs with a low Galactic extinction of $A_{V,\text{Gal}} < 0.5$ mag and avoid the Galactic plane where high extinction prevents us from having multi-filter detections of the optical-to-NIR afterglows. Overall, 130 GRBs fulfill these conditions. As most of the GROND data are not publicly accessible, we made use of published data, including those of the Gamma-ray Coordinates Networks (GCNs). Among these 130 sources, we have selected 38 bursts that have a photometric observation in at least five out of the $g'r'i'z'JHK_s$ GROND filter bands, including a minimum of one band in $JHK_s$. The photometric data of 25 GRBs in our sample were obtained from published articles, while the remaining 13 bursts were sourced from GCN circulars. To conduct our analysis, we made sure that all these bursts were observed by *Swift* and that their X-ray afterglows do not follow the form of "oddballs" according to the classification of Evans et al. (2009).

---

[1] https://www.mpe.mpg.de/~jcg/grbgen.html

For GCN data, we exclusively considered photometric observations acquired at a time during the standard late decay phase of the X-ray afterglow, where no additional spectral evolution is expected. This ensures to avoid physical effects such as optical flares or rebrightening. Our GRB sample is presented in Table A.1, where we have: the redshift, $z$, Galactic extinction, $A_{V,\text{Gal}}$, total Galactic hydrogen column density, $N_{H_{\text{tot}},\text{Gal}}$, optical/NIR bandpasses of GROND included in the GRB afterglow, the corresponding time epoch at which we will construct the SED for each burst (Sect. 4.4), the best-fit model of the GRB afterglow SED seen in referred articles and the corresponding amount of extinction $A_{V,\text{host}}$ along the GRB LoS in the host galaxy. This GRB sample contains bursts that have already undergone photometric or spectroscopic studies. This deliberate decision facilitates a direct comparison between our method and these referenced results (Sect. 4.5).

For our analysis, we constructed the SED for each burst at the time epoch given in Table A.1. The data reduction are described in Sects. 4.2–4.4. We present our final sample of dust extinguished GRBs in Sect. 4.5.

### 4.2. GROND data

When necessary, the multi-band photometry taken from GCN circulars or individual articles was corrected for the foreground Galactic extinction, using the maps of Schlafly & Finkbeiner (2011). The extinction curve of Cardelli et al. (1989) was applied with a total-to-selective extinction ratio of $R_V = 3.1$. For the cases where the data are in the Vega magnitude system, they have been converted in the AB system using the corrections from Greiner et al. (2008). We used the `ftflx2xsp` from HEASoft (version 6.31.1) to transform the photometric data into a format compatible with `xspec` for the SED analysis.

### 4.3. Swift/XRT data

The XRT light curve of each GRB afterglow was obtained from the *Swift* online repository (Evans et al. 2007). In cases where the optical-NIR and X-ray observations were non-simultaneous, we used the XRT light curve model fits to flux normalize the XRT spectra to the epoch of GROND observations (details in Appendix B). The X-ray spectra in the 0.3–10 keV energy range and in the chosen time interval were created from the automated `time-sliced spectra` tool and then normalized to the epoch of the SED. For statistical purposes and to better constrain the X-ray slopes, we have binned each spectrum by 20–25 counts per bin using the `grppha`, a sub-package of FTOOLS. This tool also associates each background and response file for the source. Only the photon counting (PC) mode was used in this analysis.

### 4.4. SED analysis

In the fireball model, a break in the synchrotron spectrum can be produced by the cooling of electrons in the GRB post-shock region (Granot & Sari 2002). The cooling break frequency, $\nu_c$, can be located between the optical ($\beta_o$) and X-ray ($\beta_X$) ranges and is expected to be $\Delta\beta = \beta_X - \beta_o = 0.5$ (Greiner et al. 2011; Zafar et al. 2011). In this case, the SED can be described as

$$F_\nu = F_0 \begin{cases} \nu^{-\beta_o} & \nu \leq \nu_c \\ \nu_c^{\beta_X - \beta_o} \nu^{-\beta_X} & \nu > \nu_c. \end{cases} \quad (6)$$

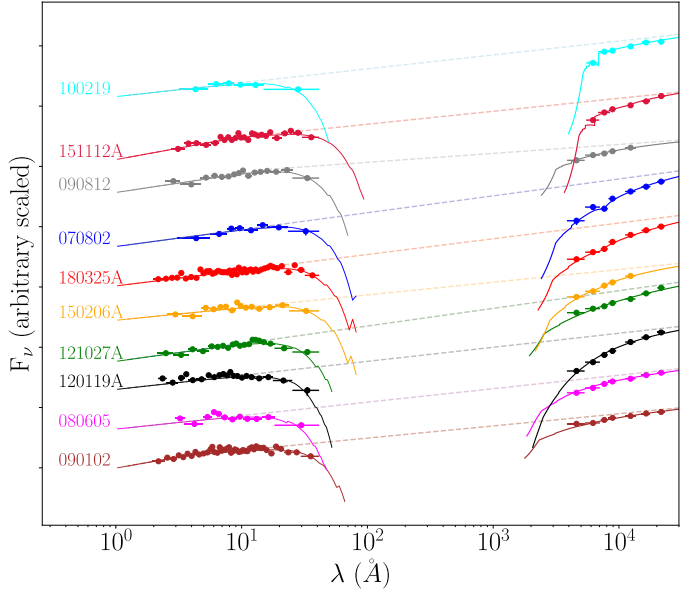

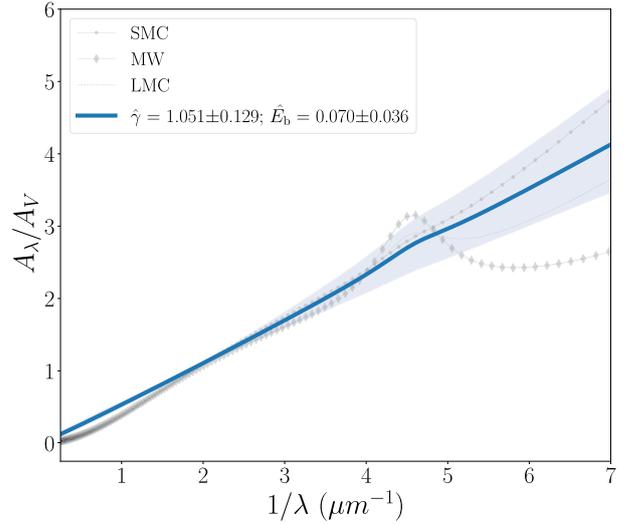

**Fig. 7.** X-ray/NIR SEDs of the 10 GRBs fitted with our simpler model of extinction, labeled on the left side and ordered by redshift. Data have been corrected for Galactic foreground extinction, the soft X-ray absorption and the host galaxy extinction (solid lines). The best fit intrinsic model to each afterglow is also indicated (dashed lines). The X-ray spectrum has been extracted at times excluding rapid fading and flares and then shifted at the GROND epoch if not available simultaneously. For better clarity, the unit of the flux density is completely arbitrary and the wavelength is plotted in the rest-frame.

**Fig. 8.** Average extinction curve of the 10 dust extinguished GRBs compared to the mean MW, LMC, and SMC laws. Shaded area represents the $1\sigma$ error.

The X-ray/NIR SED fittings were performed with xspec (v12.13.0c) (Arnaud 1996) and the best fit was found by minimizing the $\chi^2$. In the Milky Way and the host galaxy, metals are the main cause of the photoelectric absorption in the X-ray cross-section, commonly denoted as the equivalent hydrogen column density ($N_{\rm H,X}$). Values of the Galactic hydrogen column density were taken from the HI4PI survey (HI4PI Collaboration 2016). The contribution of molecular hydrogen from Willingale et al. (2013) was accounted for and we adopted the solar abundances from Wilms et al. (2000). The absorption from neutral hydrogen in the IGM was also considered (Meiksin 2006). We used both of our extinction law models, with and without the bump feature, for the host extinction. The equivalent hydrogen column density $N_{\rm H,X}$ in the host was set free. The final GRB afterglow after correcting from the Galactic components (reddening and photoelectric absorption) can be summarised as

$$F_\nu^{\rm obs} = F_\nu \times 10^{-0.4\,A_\lambda} \times \exp\left[-N_{\rm H,X}\sigma(\nu) - \tau_{\rm IGM}(\nu)\right] \quad (7)$$

where $\sigma(\nu)$ is the cross-section from Balucinska-Church & McCammon (1992) and $\tau_{\rm IGM}(\nu)$ is the optical depth of the IGM.

Based on our findings (presented in Sect. 3.2), we allowed the parameter estimations to vary within the intervals: $0 < A_V < 3$ mag, $0.2 < \gamma < 1.6$, and $0 < E_{\rm b} < 2$. A power law model of the afterglow is assumed over a broken power law when the break occurs outside the observation window ($0.002\,{\rm keV} < E_{\rm break} < 10.0\,{\rm keV}$).

### 4.5. Results

To investigate the dust extinction in GRB hosts, we focus on those with a significant amount of extinction. This choice is further motivated by our findings in Section 3.2. Of the 38 GRBs from our initial sample, we select those with a resulting extinction of $A_V > 0.3$ mag with our model of extinction which leads us to 10 GRBs (the photometric data and time interval in the X-ray for the SED fitting of our final sample are presented in Table A.3). We report in Table A.2 the best fit parameters for the X-ray/NIR SEDs of our final sample. The ten SEDs are plotted in Fig. 7 as a visual reference. Given the small number of bursts in our work, we can be subject to statistical limitations on the application of our simple extinction curve.

Seven of the final samples are best fit with a simple power law, consistent with the literature-reported results in Table A.1. Within the uncertainties, the resulting values of the visual extinction $A_V$ from our fits are also in agreement with the literature and range from $0.305 < A_V < 1.410$ mag in our case.

Figure 8 shows the average extinction curve of our 10 extinguished GRBs. The average curve is characterized by a median extinction slope $\gamma = 1.051 \pm 0.129$ and a median of the bump $E_{\rm b} = 0.070 \pm 0.036$. Given this result, its steepness corresponds to an in-between SMC-LMC curve but is closer to the SMC curve shape. This confirms the results from different studies that aim to derive the "average GRB extinction curve" or to investigate the more preferred extinction template between the three typical Local extinction curves (e.g., Kann et al. 2010; Greiner et al. 2011; Schady et al. 2012a; Covino et al. 2013; Japelj et al. 2015; Zafar et al. 2018b).

In the following, we report for each GRB the comparison of the best fit extinction curve from our method with analysis from the literature. We also include our measurement of the extinction curves of GRB 121027A and GRB 150206A.

#### 4.5.1. Comparison with spectroscopic measured GRB extinction curves

Accurate measurements of GRB host extinction curves can be derived from a spectroscopic analysis. It is mainly achieved by using the more complex model of FM86 or its modified parameterizations. We compare here our results with the measured curves found in the literature.

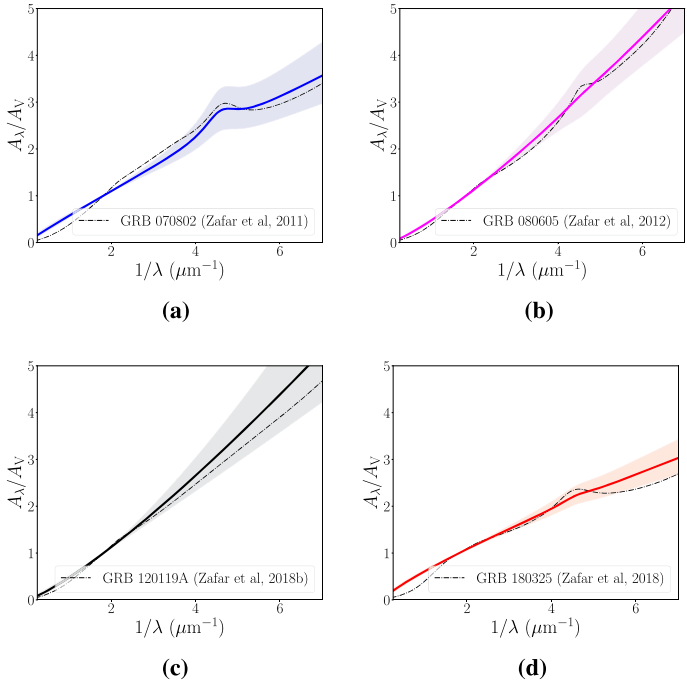

**Fig. 9.** Extinction curves of GRB host galaxies derived in this work (colored line and the corresponding 1$\sigma$ uncertainty in shaded area) compared to the measured curves with spectroscopic data in the literature (black dashed-dotted line). (a) GRB 070802, (b) GRB 080605, (c) GRB 120119A, (d) GRB 180325A.

*GRB 070802.* This is the first GRB for which the 2175 Å absorption bump was firmly identified. A first spectroscopic study with VLT/FORS2 data (Elíasdóttir et al. 2009) found that a FM90 or LMC curve provides a better fit compared to the MW or SMC one. The same result was reported in Zafar et al. (2011) using the FM86 extinction curve. This latter is plotted in Fig. 9a in comparison with our model. At a redshift $z = 2.45$, we found a similar bump with an amplitude of $E_b = 0.435 \pm 0.151$, that corresponds to the expected value for the mean LMC extinction curve (Table 1). The value of the extinction slope $\gamma = 0.938 \pm 0.135$ also confirms the steepness of the curve that is coherent with their results.

*GRB 080605.* Zafar et al. (2012) have used the FM90 parameterization to create the XRT/FORS2/GROND SED. The derived extinction curve rises steeply into the UV similar to the SMC extinction curve but presents a significant 2175 Å bump. As depicted in Fig. 9b, our analysis of the extinction curve is consistent with the shape of their measured curve within the uncertainties, yielding an extinction slope of $\gamma = 1.238 \pm 0.122$. However, as expected, the amplitude of $E_b = 0.046 \pm 0.139$ of the bump is not as prominent as in the spectroscopic study because its maximum amplitude lies outside the central wavelength of the $r$-band at $z = 1.64$.

*GRB 120119A.* Zafar et al. (2018b) showed that a FM07 parameterization similar to an SMC-like extinction curve gives a better fit to the data for this GRB. Due to its redshift, the expected 2175 Å bump that would be between the UVB and VIS arm data was not confirmed. Our best fit gives an extinction curve slightly steeper than the SMC with $\gamma = 1.239 \pm 0.170$ and the model with no bump is preferred (Fig. 9c).

*GRB 180325A.* Another detection of the 2175 bump has been reported in Zafar et al. (2018a) with XRT/X-shooter data.

Its height and area are not as prominent as the mean MW or the LMC bump. In our analysis, the best-fit model corresponds to a bump with a low amplitude of $E_b < 0.1$ and $\gamma = 0.807 \pm 0.093$ (Fig. 9d). At $z = 2.2486$, the 2175 absorption lies between the $r$ and $i$ bands, which makes its detection partially impossible in the GROND photometry.

### 4.5.2. Comparison with photometric study using the typical MW-LMC-SMC curves

We plot in Fig. 10 the four derived extinction curves of the GRBs cited below in comparison to the MW, LMC, and SMC laws.

*GRB 090102.* Using XRT/GROND with $bv$ data from UVOT, Gendre et al. 2010; Greiner et al. 2011; Schady et al. 2012a reported the GRB host galaxy extinction to be best fitted with the MW model. A MW-bump is found using our model with an amplitude of $E_b = 0.136 \pm 0.261$. Our best fit is however given by a steeper extinction slope $\gamma = 1.069 \pm 0.169$. From an idealized case (presented in Sect. 3.2) at a redshift of $z \sim 1.5$ and $A_V \sim 0.4$ mag with a MW law, we were also able to measure an overestimation of this slope by $\Delta\gamma = 0.139$ with our simple model; and this can partially explain our result for this burst. In addition, due to the presence of a break in the afterglow, having more parameters to estimate also adds constraints on the determination of the extinction slope. This case clearly shows a limit case to our simple model.

*GRB 090812 and GRB 100219.* Greiner et al. (2011), Covino et al. (2013) for GRB 090812 and Thöne et al. (2013), Bolmer et al. (2018) for GRB 100219 have reported the SMC as the preferred dust extinction model. For these GRBs, we found extinction slopes of $\gamma = 1.044 \pm 0.100$ and $\gamma = 0.983 \pm 0.096$, and no bump was identified. These values of the slope correspond to values for an in-between SMC-LMC and an SMC curve, respectively, which is agreement with the literature.

*GRB 151112A.* Bolmer et al. (2018) have found an LMC law to best fit the XRT/GROND data for this GRB. The extinction slope from our fit gives $\gamma = 1.058 \pm 0.135$ and we have also noticed the presence of a bump with $E_b = 0.239 \pm 0.052$. This slope is coherent with an in-between SMC-LMC curve, the value of the bump amplitude is however not as prominent as for a LMC curve.

### 4.5.3. New results

We present new measurements of the extinction curves for GRB 121027A where the SMC curve has been assumed in the literature and for GRB 150206A where no such study has been conducted.

*GRB 121027A.* For this ultra-long GRB (Nakauchi et al. 2013), we derive an LMC-like extinction curve with $\gamma = 0.825 \pm 0.125$ (Fig. 11a). With the best-fit power law afterglow of $\beta_X = 0.89 \pm 0.066$, we measured an extinction of $A_V = 0.938 \pm 0.061$ mag.

*GRB 150206A* is best fitted with a power law with a spectral slope of $\beta_X = 0.632 \pm 0.113$. The dust extinction in the host, as seen in Fig. 11b is characterized by a slope of $\gamma = 1.083 \pm 0.133$. At $z = 2.087$, we have measured an amplitude of the 2175 bump $E_b = 0.175 \pm 0.088$ in addition to a moderate amount of extinction $A_V = 0.780 \pm 0.094$ mag. This result confirms the presence of dust extinction stated in Selsing et al. (2019) where the X-shooter afterglow continuum presents depression in the bluer wavelengths.

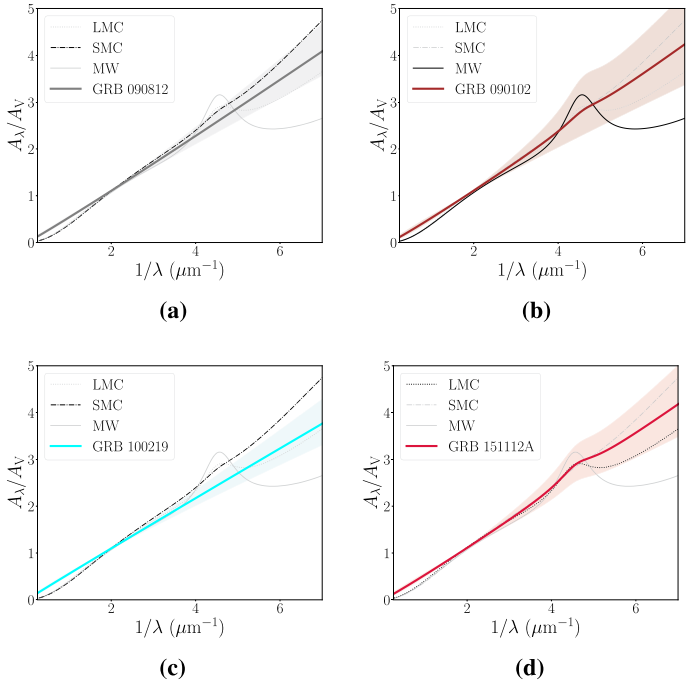

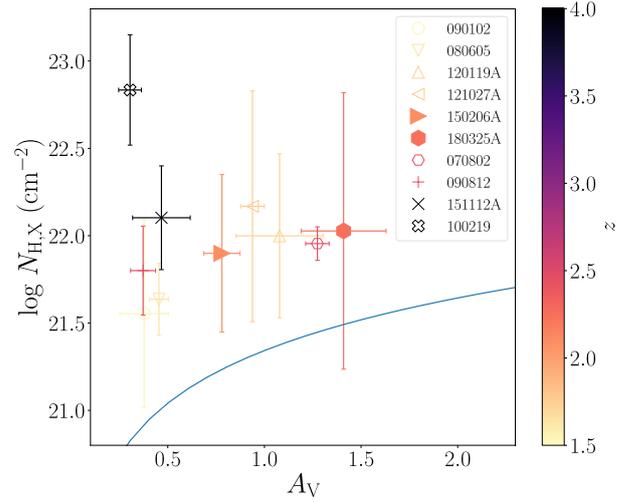

**Fig. 10.** Extinction curves of GRB host galaxies derived in this work (colored line and the corresponding $1\sigma$ uncertainty in shaded area) compared to three mean extinction curves, black lines represent the best fit in the literature. (a) GRB 090812, (b) GRB 090102, (c) GRB 100219, (d) GRB 151112A.

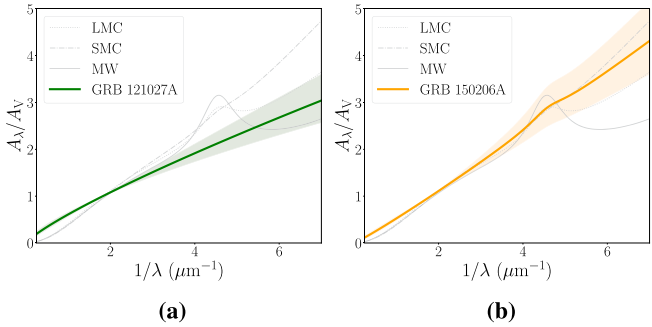

**Fig. 11.** Measured extinction curves for GRB 121027A (left) and GRB 150206A (right) compared to the MW, SMC and LMC extinction curve. The shaded areas correspond to the $1\sigma$ uncertainty. (a) GRB 121027A, (b) GRB 150206A.

#### 4.5.4. Equivalent hydrogen column density

Our measurements of the equivalent hydrogen densities $N_{H,X}$ are reported in Table A.2. Our results are in agreement with the literature within $2\sigma$ error. From Fig. 12, we notice an increase of the column density with the redshift. One possible reason for this trend is simply the increase of the intergalactic medium absorption at larger distances (Campana et al. 2010, 2012, 2015; Starling et al. 2013). A noticeable lack of low $N_{H,X}$ values may also causes a bias towards GRB sources at higher redshifts. Using a complete sample of GRBs, Covino et al. (2013) investigated $N_{H,X}$ as a function of $A_V$ and affirmed that definitive conclusions cannot be derived from such method. We do not discuss further the resulting hydrogen column density here as more refined models can be used to fit XRT and to derive $N_{H,X}$ (Buchner et al. 2017). In comparison to the redshift dependence, they find that the distribution of $N_{H,X}$ follows a random distributions of sources in an axisymmetric ellipsoid of gas cloud.

**Fig. 12.** Distribution of $N_{H,X}$ vs. $A_V$ for the 10 studied GRBs. The blue line represents the relation between the dust-to-metals relation in the Galaxy: $N_{H,X}/A_V = 2.2^{+0.4}_{-0.3} \times 10^{21}$ cm$^{-2}$ mag$^{-1}$ (Watson 2011). The filled symbols to the newly analysed GRBs.

## 5. Discussion

### 5.1. Extinction curve and dust properties traced with $\gamma$

We have shown that with only the extinction slope $\gamma$ parameter, we shown evidence of the diversity of extinctions along the different LoS that we have observed in GRB host galaxies. Even with given a small number of bursts, a visible trend in our results can be reported. We can see in Fig. 13, $\gamma$ decreases as the extinction $A_V$ increases.

Similarly, Salim & Narayanan (2020) have examined direct observations of various MW and LMC/SMC sightlines including the 30 Doradus region and defined the overall UV-optical slope, $A_{1500}/A_V$, as a function of $A_V$. Analyzing this ratio has allowed them to distinguish between the different shapes of these extinction curves. They confirm that not only the extinction curves from LMC and SMC sightlines are steeper, but they are also less extinguished and present a smaller strength of the 2175 Å bump than for typical MW sightlines; this corroborates the trend highlighted in our result. In addition, it is not rare to observe instances of convergence among extinction curves. From the Gordon et al. (2003) sample, four extinction curves in the LMC are found to be indistinguishable from MW curves: same values of $A_V$ with similar slopes. The inverse case has been observed (HD 210121, Larson et al. 2000, HD 164340, Clayton et al. 2000), where the MW sightlines are almost as steep as the SMC with no-to-a slight bump. Having this diversity of extinction curves (even in our Galaxy) favors the use of a simplistic model that can describe the whole continuum of possible laws with solely the use of the extinction slope, $\gamma$.

It is known that the difference of extinction curves in galaxies is mainly correlated to the variation in the grain-size distribution. Shallower curves are expected in rich environments of larger grains. This is demonstrated by investigating the evolution of extinction curves in galaxies which provides us with an understanding of the dust grain properties. The grain growth, which is the accretion of gas-phase metals onto the surface of pre-existing grains, can lead to variations in the grain-size distribution (Asano et al. 2014) and, consequently, in the shape of extinction curves. In the early stage of galaxy evolution, larger grains originating from stars are the main cause of flatter curves. Afterward, shattering causes an abundance of small grains and the 2175 Å bump becomes more evident, with a steeper extinction curve due to the grain growth.

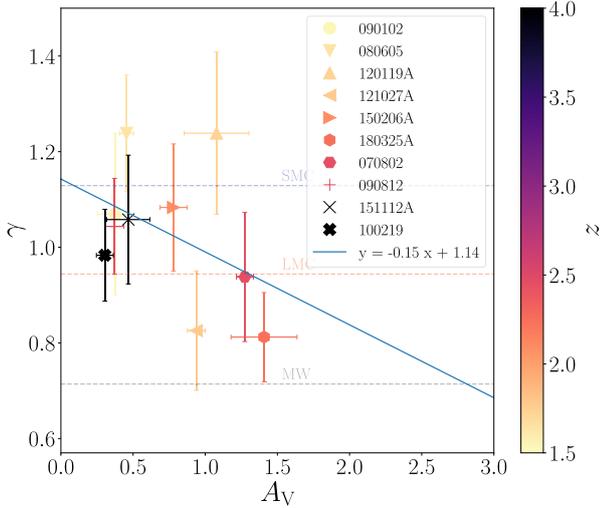

**Fig. 13.** Distribution of $\gamma$ vs $A_V$ for the 10 studies GRBs. Data points are color-coded according to the redshift. Dashed lines correspond to the expected values $\gamma$ for the average MW, LMC, and SMC laws (Table 1). The blue line illustrates the best fit using an ordinary least-squares linear regression.

In addition, dust plays a crucial role in affecting and reshaping the emission of galaxies by absorbing and scattering the stellar light. The method most commonly adopted to model this effect is to introduce a simple effective law that encapsulates the complex blending of dust properties and geometrical distributions of stars and dust. This law is called attenuation curve to avoid any confusion with the extinction curve, which samples a single LoS. Corre et al. (2018) compared literature reports of measured extinction curves with their measured effective attenuation curves in galaxies hosting a GRB with radiative transfer simulations from Seon & Draine (2016). These comparisons illustrate that various dust-stars geometry and a clumpy ISM could reproduce the diversity of curves. However, the models were limited to the three extinction curves MW, LMC, and SMC. The measurement of the slope of the extinction curve opens the way to define dust models able to reproduce the distribution of $\gamma$ values and to be used in radiation transfer simulations in order to get a more precise description of the dusty ISM of the host galaxies (e.g., Liu & Hirashita 2019; Lin et al. 2021; Hsu et al. 2023).

### 5.2. SVOM/COLIBRI synergy

The Space-based multi-band astronomical Variable Objects Monitor mission (SVOM, Gonzalez & Yu 2018) will provide a versatile satellite equipped with multi-wavelength instruments (ECLAIRs, Gamma Ray Monitor (GRM), Microchannel X-ray Telescope (MXT), and Visible Telescope (VT)). On June 22, 2024, the SVOM satellite was launched and injected into a low Earth orbit (~600 km) with an anti-solar pointing strategy and an inclination angle of ~30°. This so-called B1-law attitude of SVOM has been chosen to optimize both the time spent in observing at high galactic latitudes and the GRB detection rate during the night side. The SVOM satellite pointing strategy has been thought to maximize the discovery potential of the ground-based telescopes operating at optical and NIR wavelengths. These SVOM space and ground-based facility synergies are consolidated by the building of a dedicated and coordinated follow-up segment with robotic telescopes including the 1.3 m COLIBRI (Basa et al. 2022) (in Mexico) and the C-GFT 1.2 m (in China) telescopes. In particular, COLIBRI will be able to observed the sky with the DDRAGO two channels camera (*gri* and *zy* filters) and the CAGIRE infrared camera (*JH* filters). Despite the absence of the *K*-band, the COLIBRI instrumental setup is very similar to GROND. Hence, the association of the SVOM instruments with COLIBRI will allow us to study the color evolution of GRB afterglows and characterize the dust properties from their host galaxies during the next decade. In this perspective, an end-to-end simulation can be addressed in the meantime of the first X-ray-to-NIR data from SVOM/COLIBRI.

## 6. Summary and conclusion

We present a simple model of dust extinction based on a power law form and designed for photometric studies. The singularity of our model is its independence from prior assumptions on the size and distribution of the dust. With this simple model, we aim to characterize the dust properties in GRB host galaxies with only two parameters: the extinction slope, $\gamma$, and the 2175 Å absorption amplitude, $E_b$.

To validate our model, we first simulate "perfect" GRB afterglows based on a selection of fixed extinction templates and using the seven-band filters of GROND. As our model is designed for simplicity, we acknowledge certain limitations that we aim to address before testing with real data. The goal at this stage was first, to estimate the intrinsic behavior of our proposed model under idealized conditions and, second, to refine and validate the analysis that was later applied to selected *Swift*/GROND GRBs. Our model succeeds in determining the value of the extinction $A_V$ except for extreme cases of low redshift and high extinction for afterglows presenting the 2175 Å absorption bump. This is mainly due to the lack of UV coverage in such photometric studies. As a consequence, we have an underestimation of $A_V$ correlated to an overestimation of $\gamma$. We highlight also that in photometric studies like ours, the detection and the measurement of the bump depend on the quantity of extinction and the redshift.

Second, we have performed broadband X-ray-to-NIR SED fitting on a sample of *Swift*/GROND GRBs using our model. For data where no quasi-simultaneous observations are available, we have normalized the selected time-interval *Swift* spectra to the GROND reference time. The selection criteria for our sample and bursts to be analyzed are mainly based on the results obtained in the simulation part of the study. The introduction of the extinction slope, $\gamma$, is indeed a good approach for the measurement of the absolute extinction of GRB host, compared to the $R_V$-dependent fixed laws of the MW, LMC, and SMC. In accordance with previous studies, the average extinction curve determined in this work exhibits a similar behavior to an intermediate between the SMC and LMC-like extinction curve. The fact that our model correctly fits the GRB afterglows and that there are no physical reasons to choose a fixed extinction template to describe the GRB host are all points in favor of our simple model for such applications.

This study also serves as a preparatory work for the scientific exploitation of the upcoming SVOM mission and the COLIBRI telescope.

*Acknowledgements.* NR acknowledges D. Götz for his helpful comments. NR acknowledges support by the Centre National d'Etudes Spatiales for the funding of his PhD. We are also grateful to Agnes Monod-Gayraud for her reviewing of the manuscript and assistance with the English corrections, and the anonymous referee for providing comments that helped improve the quality of this work. This work - made use of data supplied by the UK *Swift* Science Data Centre at the University of Leicester. – received support from the French government under the France 2030 investment plan, as part of the Initiative d'Excellence d'Aix-Marseille Université – A*MIDEX – AMX-19-IET-008.

## Appendix A: Additional tables

**Table A.1.** The 38 GRBs analyzed in this paper.

| GRB | $z$ | $A_{V,Gal}$ (mag) | $N_{H_{tot},Gal}$ ($10^{21}$ cm$^{-2}$) | Optical/NIR bandpasses | Epoch (h) | best-fit model afterglow/ext. curve | $A_{V,host}$ (mag) |
|---|---|---|---|---|---|---|---|
| 150301B | 1.517 | 0.210 | 0.640 | $g'r'i'z'JH$ | 5.000 | - | - |
| 090102 | 1.547 | 0.125 | 0.463 | $g'r'i'z'JHK_s$ | 3.090 | bknpow/MW[1] | 0.41±0.03 |
| 080605 | 1.640 | 0.365 | 1.030 | $g'r'i'z'JHK_s$ | 2.785 | pow/FM90[2] | $0.50^{+0.12}_{-0.10}$ |
| 191011A | 1.722 | 0.043 | 0.143 | $g'r'i'z'JHK_s$ | 0.100 | SMC[3] | 0.43±0.03 |
| 120119A | 1.728 | 0.296 | 1.035 | $g'r'i'z'JHK_s$ | 0.058 | pow/FM07[4] | 1.02±0.11 |
| 121027A | 1.773 | 0.053 | 0.162 | $g'r'i'z'JH$ | 49.200 | pow/SMC[5] | 1.1 |
| 080514B | 1.800 | 0.151 | 0.425 | $g'r'i'z'JHK_s$ | 21.768 | SMC†[6] | ≤0.14 |
| 170113A | 1.968 | 0.304 | 1.079 | $g'r'i'z'JH$ | 372.004 | SMC†[7] | $0.00^{+0.00}_{-0.00}$ |
| 081230 | 2.000 | 0.038 | 0.155 | $g'r'i'z'JHK_s$ | 5.44 | SMC[8] | $0.22^{+0.14}_{-0.08}$ |
| 150206A | 2.087 | 0.042 | 0.185 | $g'r'i'z'J$ | 10.300 | - | - |
| 180325A | 2.248 | 0.046 | 0.126 | $g'r'i'z'JHK_s$ | 1.626 | pow/FM90[9] | $1.58^{+0.10}_{-0.12}$ |
| 121024A | 2.298 | 0.272 | 0.702 | $g'r'i'z'JHK_s$ | 29.722 | pow/SMC[10] | 0.18±0.04 |
| 120815A | 2.358 | 0.308 | 1.168 | $g'r'i'z'JHK_s$ | 2.163 | bknpow/SMC[11] | $0.32^{+0.02}_{-0.02}$ |
| 070802 | 2.450 | 0.071 | 0.265 | $g'r'i'z'JHK_s$ | 0.708 | pow/FM86[12] | 0.8-1.5 |
| 090812 | 2.452 | 0.064 | 0.213 | $g'r'i'z'J$ | 2.183 | bknpow/SMC[13] | $0.23^{+0.08}_{-0.06}$ |
| 081121 | 2.512 | 0.136 | 0.420 | $g'r'i'z'JHK_s$ | 3.945 | bknpow/SMC[1] | $0.07^{+0.01}_{-0.01}$ |
| 081118 | 2.580 | 0.118 | 0.396 | $g'r'i'z'JHK_s$ | 10.93 | - | - |
| 090426 | 2.609 | 0.045 | 0.150 | $g'r'i'z'JH$ | 12.852 | pow/SMC†[14] | 0.0 |
| 080210 | 2.641 | 0.217 | 0.693 | $g'r'i'z'JHK_s$ | 1.001 | bknpow/SMC[15] | 0.18±0.03 |
| 161023A | 2.708 | 0.091 | 0.291 | $g'r'i'z'JHK_s$ | 27.249 | pow/SMC[16] | 0.09±0.03 |
| 110731A | 2.830 | 0.480 | 1.510 | $g'r'i'z'JH$ | 65.7696 | pow/SMC[17] | 0.24±0.06 |
| 120404A | 2.876 | 0.131 | 0.392 | $g'r'i'z'JH$ | 19.451 | pow/MW[18] | 0.22±0.05 |
| 081028 | 3.038 | 0.093 | 0.452 | $g'r'i'z'J$ | 31.300 | SMC[19] | <0.22 |
| 120922A | 3.100 | 0.396 | 1.152 | $g'r'i'z'JHK_s$ | 1.500 | - | - |
| 081228 | 3.440 | 0.431 | 1.269 | $r'i'z'JHK_s$ | 0.719 | pow/SMC[20] | $0.12^{+0.06}_{-0.08}$ |
| 160203A | 3.520 | 0.187 | 0.725 | $g'r'i'z'JHK_s$ | 0.154 | FM90[21] | <0.10 |
| 081029 | 3.848 | 0.082 | 0.293 | $g'r'i'z'JHK_s$ | 5.823 | pow/SMC†[22] | $0.03^{+0.02}_{-0.03}$ |
| 131117A | 4.042 | 0.050 | 0.134 | $g'r'i'z'JHK_s$ | 0.102 | bknpow/FM90[4] | <0.11 |
| 090516 | 4.109 | 0.134 | 0.536 | $g'r'i'z'JHK_s$ | 14.803 | pow/SMC[23] | $0.19^{+0.03}_{-0.03}$ |
| 120712A | 4.175 | 0.112 | 0.389 | $g'r'i'z'JHK_s$ | 9.582 | bknpow/SMC[23] | $0.08^{+0.03}_{-0.08}$ |
| 140614A | 4.233 | 0.330 | 1.148 | $r'i'z'JHK_s$ | 0.647 | bknpow/SMC[23] | $0.11^{+0.17}_{-0.05}$ |
| 151112A | 4.270 | 0.036 | 0.149 | $r'i'z'JHK_s$ | 11.397 | bknpow/LMC[23] | $0.50^{+0.21}_{-0.11}$ |
| 090205A | 4.650 | 0.312 | 1.012 | $r'i'z'JHK_s$ | 6.485 | pow/LMC[23] | $0.14^{+0.04}_{-0.05}$ |
| 100219A | 4.667 | 0.203 | 0.791 | $r'i'z'JHK_s$ | 9.513 | pow/SMC[24] | 0.24±0.06 |
| 140428A | 4.680 | 0.026 | 0.095 | $r'i'z'JHK_s$ | 2.601 | bknpow/SMC[23] | $0.30^{+0.32}_{-0.23}$ |
| 140311A | 4.954 | 0.099 | 0.235 | $r'i'z'JHK_s$ | 9.339 | pow/SMC[23] | $0.07^{+0.03}_{-0.03}$ |
| 111008A | 4.990 | 0.014 | 0.084 | $r'i'z'JHK_s$ | 6.848 | bknpow/FM90[4] | 0.12±0.04 |
| 130606A | 5.931 | 0.064 | 0.214 | $r'i'z'JHK_s$ | 8.370 | pow/FM90[4] | <0.07 |

**Notes:** Columns 7-8 are the best-fit models found in the literature for each GRB with the corresponding total amount of extinction in the V-band. When different sources are found, spectroscopic studies are preferred. For analysis with only optical data, the shape of the SED, whether fit with a broken power law or a power law, is not indicated. † SMC law was assumed since the different extinction laws could not be distinguished. [1]Schady et al. (2012a); [2]Zafar et al. (2012); [3]Heintz et al. (2023)[20]; [4]Zafar et al. (2018b); [5]Nakauchi et al. (2013); [6]Rossi et al. (2008); [7]Gompertz et al. (2018); [7]Krühler et al. (2011); [9]Zafar et al. (2018a); [10]Varela et al. (2016); [11]Japelj et al. (2015); [12]Zafar et al. (2011); [13]Covino et al. (2013); [14]Nicuesa Guelbenzu et al. (2011); [15]De Cia et al. (2011); [16]de Ugarte Postigo et al. (2018); [17]Ackermann et al. (2013); [18]Guidorzi et al. (2014); [19]Margutti et al. (2010); [20]Greiner et al. (2011); [21]Heintz et al. (2019); [22]Nardini et al. (2011); [23]Bolmer et al. (2018); [24]Thöne et al. (2013)

**Table A.2.** Dust extinguished GRBs analyzed in this paper.

| GRB | z | Afterglow | $N_{H,X}$ ($10^{22}$ cm$^{-2}$) | $A_V$ (mag) | $\gamma$ | $E_b$ | $\beta_X$ | $\beta_o$ | Break (keV) | $\chi^2$ (d.o.f.) |
|---|---|---|---|---|---|---|---|---|---|---|
| GRB090102 | 1.547 | bknpow | 0.358±0.104 | 0.378±0.125 | 1.069±0.169 | 0.136±0.261 | 1.105 | 0.605±0.032 | 3.352±0.0 | 2.303 (44) |
| GRB080605 | 1.64  | pow    | 0.432±0.268 | 0.454±0.048 | 1.238±0.122 | 0.046±0.139 | 0.671±0.056 | -           | -           | 2.178 (13) |
| GRB120119A | 1.728 | pow    | 0.998±0.339 | 1.078±0.225 | 1.239±0.170 | -           | 0.699±0.029 | -           | -           | 1.857 (23) |
| GRB121027A | 1.773 | pow    | 1.473±0.322 | 0.938±0.061 | 0.825±0.125 | -           | 0.89±0.066  | -           | -           | 1.494 (22) |
| GRB150206A | 2.087 | pow    | 0.793±0.281 | 0.780±0.094 | 1.083±0.133 | 0.175±0.088 | 0.632±0.113 | -           | -           | 1.402 (14) |
| GRB180325A | 2.248 | pow    | 1.064±0.172 | 1.410±0.220 | 0.807±0.093 | 0.095±0.020 | 0.786±0.029 | -           | -           | 2.802 (45) |
| GRB070802 | 2.450 | pow    | 0.901±0.722 | 1.274±0.061 | 0.938±0.135 | 0.435±0.151 | 0.840±0.075 | -           | -           | 2.440 (9)  |
| GRB090812 | 2.452 | bknpow | 0.631±0.351 | 0.372±0.0646 | 1.044±0.100 | -          | 0.943       | 0.443±0.100 | 0.716±0.599 | 0.907 (15) |
| GRB151112A | 4.27 | bknpow | 1.267±0.639 | 0.467±0.149 | 1.058±0.135 | 0.239±0.052 | 1.143       | 0.643±0.124 | 1.644±0.569 | 1.217 (24) |
| GRB100219 | 4.667 | pow    | 6.824±3.299 | 0.305±0.059 | 0.983±0.096 | -           | 0.691±0.095 | -           | -           | 2.133 (7)  |

**Notes.** Column (3) is the best fit intrinsic afterglow (power law or broken power law). Columns (3) to (10) are the host galaxy equivalent column density $N_{H,X}$, the host galaxy extinction $A_V$, the extinction slope $\gamma$, the amplitude of the 2175 Å absorption which is not mentioned when the extinction model without the Drude profile gives the better fit, the optical/X-ray spectral slopes following $\beta = \beta_X - \beta_o = 0.5$ (bknpow) or $\beta_X = \beta_o$ (pow), and the cooling break energy. The last column gives the reduced $\chi^2_{red}$ and degrees of freedom (d.o.f.).

**Table A.3.** GROND photometry and the X-ray time interval used for the SED fitting of the final 10 GRBs.

| GRB | $g'$ | $r'$ | $i'$ | $z'$ | $J$ | $H$ | $K_s$ | X-ray time interval $\Delta t^c$ (h) |
|---|---|---|---|---|---|---|---|---|
| 090102[1] | 20.98±0.03[b] | 20.64±0.03[b] | 20.03±0.03[b] | 19.64±0.03[b] | 19.03±0.04[b] | 18.25±0.07[b] | 17.59±0.07[b] | 1.48-7.26 |
| 080605[2] | 20.76±0.05 | 20.15±0.05 | 19.66±0.05 | 19.35±0.05 | 19.01±0.09 | 18.56±0.09 | 18.26±0.11 | 1.67-5.56 |
| 120119A[4,d] | 18.8±0.1 | 17.7±0.1 | 16.8±0.1 | 16.2±0.1 | 15.2±0.1 | 14.6±0.1 | 14.0±0.1 | 0.39-2.5 |
| 121027A[5,d] | 22.1±0.2 | 21.6±0.1 | 21.3±0.1 | 20.9±0.1 | 20.4±0.1 | 19.7±0.1 | 19.7±0.1 | 62.5-152.78 |
| 150206A[6,d] | 22.6 0.2 | 21.9 0.1 | 21.2 0.1 | 20.8 0.1 | 20.0 0.2 | >20.2 | >19.4 | 9.27-11.33 |
| 180325A[7] | 20.44±0.04 | 19.56±0.03 | 19.16±0.03 | 18.29±0.03 | 17.42±0.05 | 16.84±0.05 | 16.38±0.07 | 1.32-1.93 |
| 070802[8] | 23.16±0.42[a] | 21.46±0.16[a] | 21.66±0.1[a] | 20.41±0.15[a] | 19.52±0.1[a] | 18.92±0.12[a] | 18.29±0.13[a] | 1.11-119.44 |
| 090812[9,d] | 22.15±0.15 | 21.50±0.08 | 21.30±0.13 | 21.02±0.12 | 20.57±0.15 | >19.8 | >18.9 | 2.78-69.44 |
| 151112A[10] | >25.4 | 23.58±0.09 | 22.63±0.08 | 22.23±0.09 | 21.74±0.26 | 21.25±0.22 | 20.59±0.26 | 5.67-10.75 |
| 100219A[10] | >24.2 | 22.77±0.09 | 21.37±0.06 | 21.24±0.08 | 20.72±0.20 | 20.23±0.19 | 20.12±0.34 | 2.78-8.33 |

**Notes:** Magnitudes are in the AB system unless otherwise indicated. [a]Magnitudes already corrected for the Galactic foreground reddening. [b]Vega magnitude system. [c]Time interval at which the X-ray spectrum has been created and used for the broad-band SED. [d]Data from GCN circulars. [1]Gendre et al. (2010); [2]Zafar et al. (2012); [4]Schady et al. (2012b); [5]Sudilovsky et al. (2012); [6]Schweyer et al. (2015); [8]Krühler et al. (2008); [9]Updike et al. (2009); [10]Bolmer et al. (2018)

## Appendix B: Normalizing X-ray spectra to the SED epoch

To create the instantaneous X-ray-to-NIR SED, we must normalize the X-ray spectra to the time epoch of the optical observation. First, we have selected for each GRB a period in the X-ray light curve during the late normal decay phase. This is to ensure that no spectral evolution (flares or plateau phase) that could influence the intrinsic afterglow slope is present (except for GRB 100219 where we have poor statistics in the later phase). This temporal decay is fitted using either a simple power law or a smoothed broken power law using the equations (Zaninoni et al. 2013):

$$F(t) = \begin{cases} N\, t^{-\alpha} \\ N\left(\left(\frac{t}{t_b}\right)^{-\frac{\alpha_1}{s}} + \left(\frac{t}{t_b}\right)^{-\frac{\alpha_2}{s}}\right)^s \end{cases} \quad (B.1)$$

where $N$ is the normalization factor, $\alpha_1$ and $\alpha_2$ are the temporal slopes, $t_b$ is the time of the break, and the smoothness parameter, $s$, is fixed to -0.3.

A time interval is then selected for the X-ray spectrum that is created in the SED analysis. The mean flux is calculated by integrating the total flux in the selected time interval, $\Delta t$, with

$$F_m = \frac{\int_{t_1}^{t_2} F_\nu\, dt}{\Delta t} \quad (B.2)$$

The ratio between this mean flux and the X-ray flux at the optical time reference is used to normalize the X-ray spectrum for the creation of the quasi-simultaneous broadband SED.

The X-ray-to-NIR light curves used for the creation of the SEDs of our sample are shown in Fig. B.1-B.10. For each GRB, best-fit models and the spectral slope parameters of XRT data are shown in red line. The grey area represents the chosen time interval at which the X-ray spectrum was used for the broadband SED fitting at the optical reference time. For GRB 100219 and GRB 070802, we have chosen X-ray time interval during the plateau phase and the late standard phase respectively because of the poor statistics at the observation time. For GRB 120119A, extrapolation of the PC mode light curve has confirmed that the GROND time epoch still lies on the same spectral evolution.

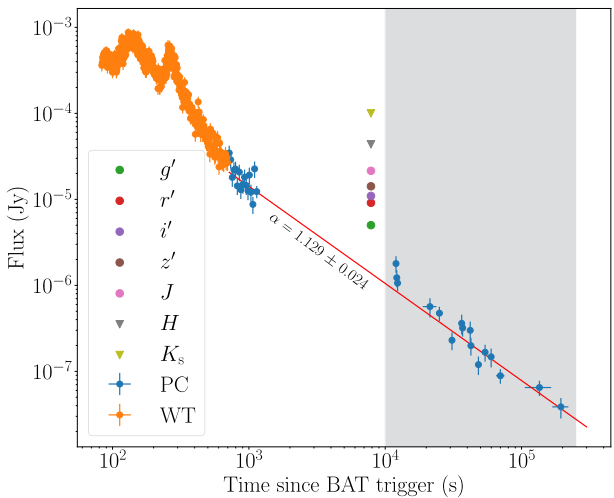

**Fig. B.1.** GRB 090812

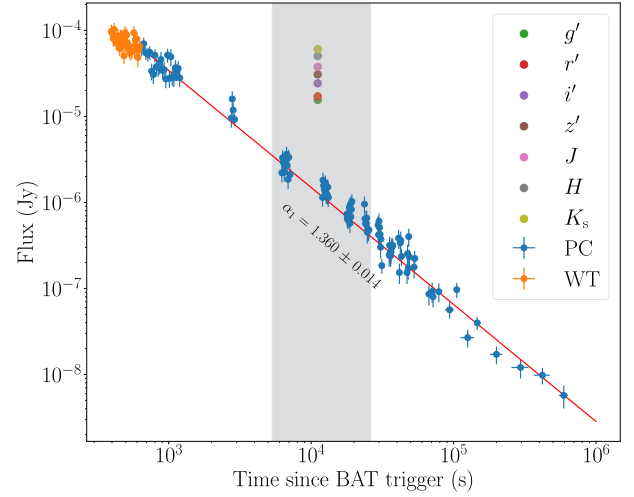

**Fig. B.2.** GRB 090102

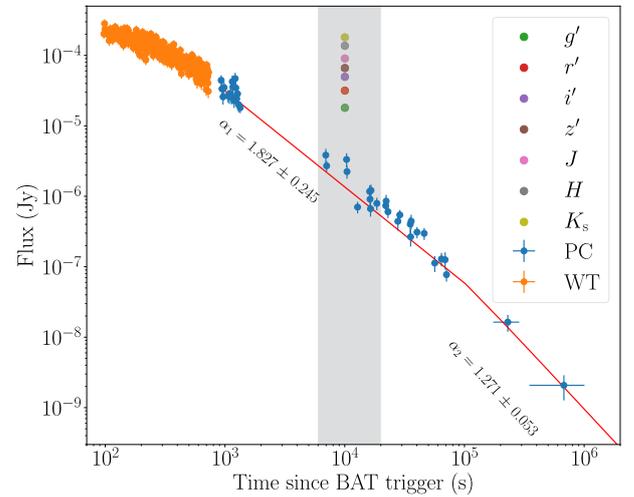

**Fig. B.3.** GRB 080605

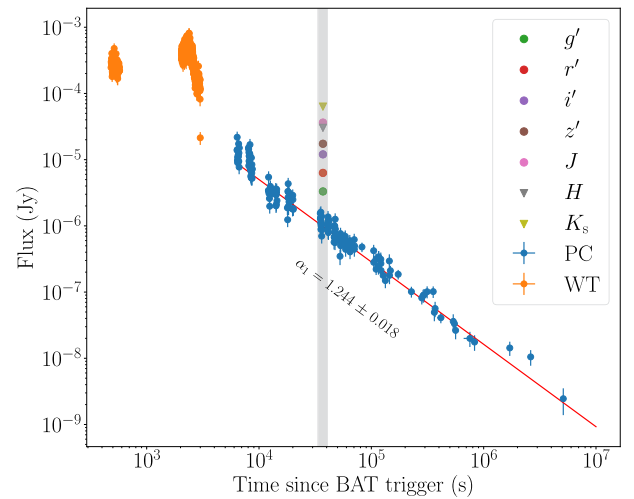

**Fig. B.4.** GRB 150206A

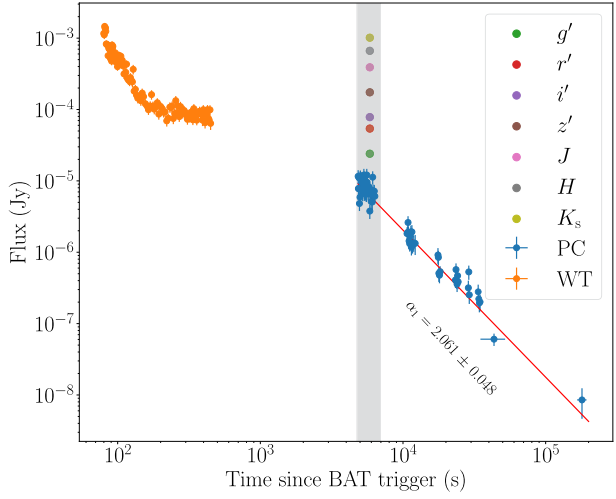

**Fig. B.5.** GRB 180325A

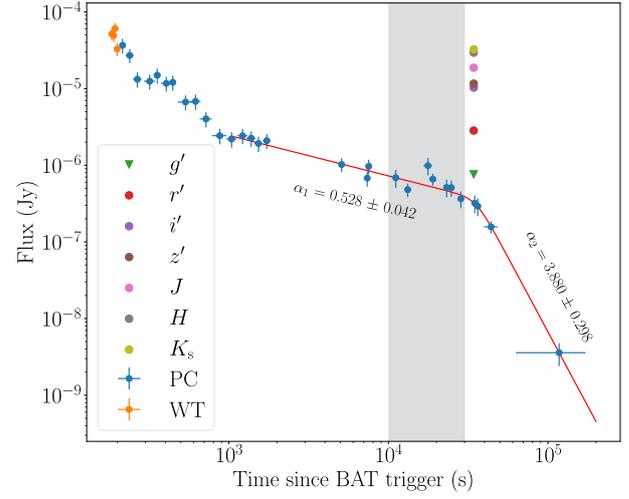

**Fig. B.8.** GRB 100219

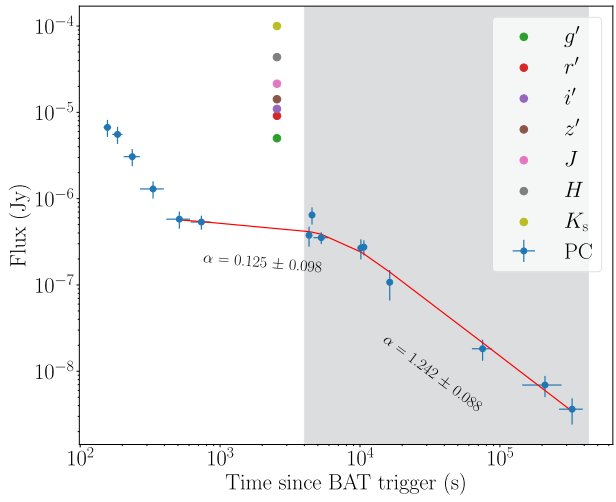

**Fig. B.6.** GRB 070802

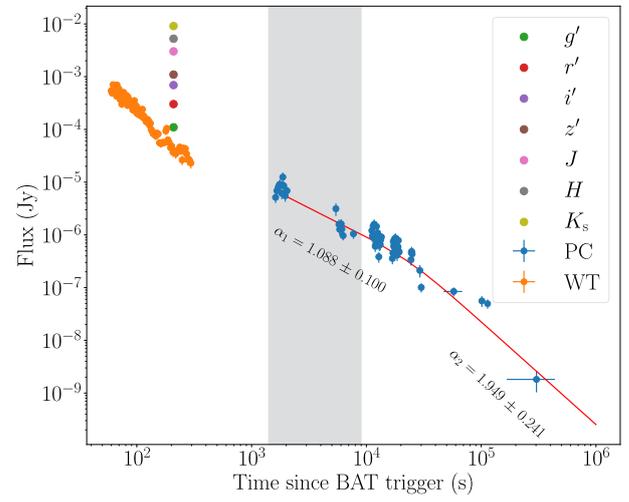

**Fig. B.9.** GRB 120119A

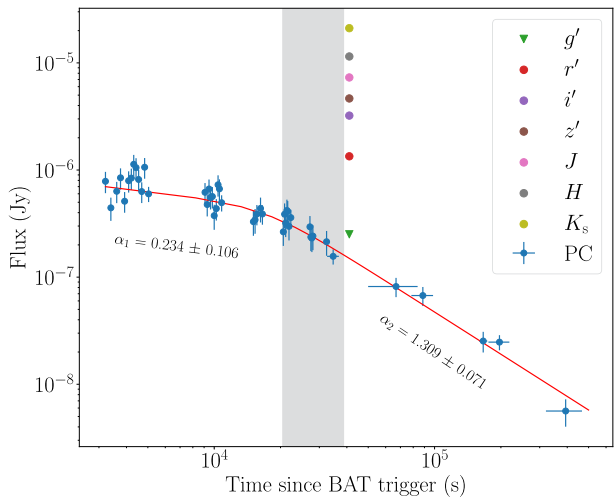

**Fig. B.7.** GRB 151112A

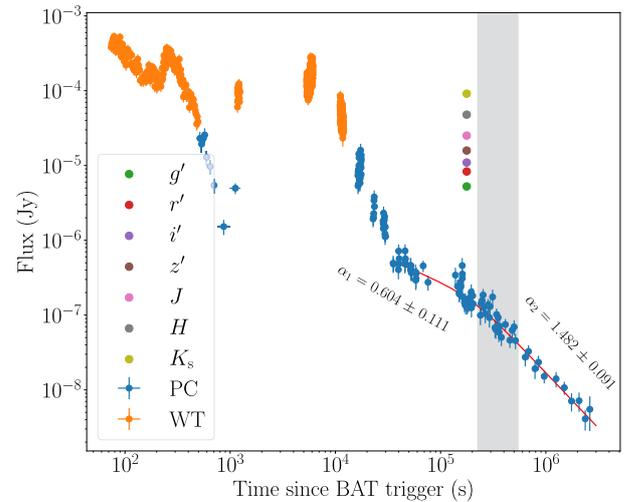

**Fig. B.10.** GRB 121027A